\documentstyle[11pt]{article}

\newcommand{\bigzerou}{%
\smash{\lower1.7ex\hbox{\bg 0}}}
\renewcommand{\theequation}{\arabic{section}.\arabic{equation}}
\newcommand{\beq}{\begin{equation}}
\newcommand{\enq}{\end{equation}}

\newcommand{\mapright}[1]{%
\smash{\mathop{%
\hbox to 1.0cm{\rightarrowfill}}\limits^{#1}}}
\newcommand{\mapleft}[1]{%
\smash{\mathop{%
\hbox to 1.3cm{\leftarrowfill}}\limits^{#1}}}

\newcommand{\no}{\nonumber}

\newtheorem{thm}{Theorem}

\newcommand{\beqy}{\begin{eqnarray}}
\newcommand{\enqy}{\end{eqnarray}}

\begin{document}


\begin{titlepage}
\vglue 3cm

\begin{center}
\vglue 0.5cm
{\Large\bf Hecke Operator and $S$-Duality of ${\cal N}=4$ Super Yang-Mills for $ADE$ Gauge Group
on $K3$} 
\vglue 1cm
{\large Toru Sasaki} 
\vglue 0.5cm
{\it Department of Physics,
Hokkaido University, Sapporo 060-0810, Japan}\\
{ sasaki@particle.sci.hokudai.ac.jp}

\baselineskip=12pt

\vglue 1cm
\begin{abstract}
 We determine the partition functions
 of ${\cal N}=4$ super Yang-Mills gauge theory 
 for some $ADE$ gauge groups on $K3$,
 under the assumption that they are holomorphic.
 Our partition functions satisfy the gap condition
 and Montonen-Olive duality at the same time,
 like the $SU(N)$ partition functions of Vafa and Witten.
 As a result we find a close relation between
 Hecke operator and $S$-duality of ${\cal N}=4$
  super Yang-Mills for $ADE$ gauge group
 on $K3$.
\end{abstract}
\end{center}
\end{titlepage}
\section{Introduction}
\label{sec:1}
\setcounter{equation}{0}
 
This article provides a new result for our trial to determine the ${\cal N}=4~ADE$
partition functions on $K3$.
The difficulty in this trial was due to
the compatibility between the Montonen-Olive duality
and the gap condition \cite{jin3}.
In this article, we define the deformed Hecke operator
(a kind of Hecke operator) as a new tool.
By using this operator, we overcome this difficulty
and determine the ${\cal N}=4~ADE$ partition functions
on $K3$ for some groups,
which satisfy the  Montonen-Olive duality
and the gap condition at the same time.

Vafa and Witten have first investigated the twisted ${\cal N}=4$
exact partition functions for $SU(N)$ on some K\"ahler 4-manifolds,
such as $K3,T^4,C{\bf P}^2,C{\hat {\bf P}}^2,ALE$ \cite{vafa-witten}.
The twisted ${\cal N}=4$ super Yang-Mills theory 
is a kind of topological field theory \cite{laba}
and its partition functions are expected to have simple form.
In fact the partition function is 
the summation of the Euler number of the moduli space of ASD equations.
The moduli space which appears in the twisted $N=4$ super Yang-Mills theory
is that of the irreducible connections.
The dimension of this moduli space is given by
the Atiyah-Hitchin-Singer formula \cite{AHS}.
This formula shows 
that the moduli space of small instanton number  vanishes.
Thus the corresponding partition functions have a gap \cite{vafa-witten}.
This condition is called the gap condition.
On the other hand, the  twisted ${\cal N}=4$ super Yang-Mills theory
must obey $S$-duality of ${\cal N}=4$ super Yang-Mills theory.
$S$-duality of ${\cal N}=4$ super Yang-Mills theory is believed to be the one 
considered by Montonen and Olive \cite{M-O}.
In the Montonen-Olive duality \cite{M-O},
the exchange of strong/weak coupling constant 
accompanies 
the exchange of the gauge groups ${\cal G}\leftrightarrow {\hat {\cal G}}$.
When ${\cal G}$ is $ADE$ gauge group,
 ${\hat {\cal G}}={\cal G}/\Gamma_{\cal G}$. 
Here $\Gamma_{\cal G}$ is center of ${\cal G}$
 and $|\Gamma_{\cal G}|$ is the number of elements of $\Gamma_{\cal G}$.
The above general structure of the twisted ${\cal N}=4$
partition functions are called Vafa-Witten conjecture.

After the Vafa-Witten conjecture was proposed,
the results satisfying this appeared in many examples \cite{ E-S,iqbal,jin,jin2,bonelli,lozano,m-n,m-v,yoshi}.
What about the case of $ADE$ gauge groups ?
Indeed, there was a conjecture for $ADE$ gauge groups
supporting the Montonen-Olive duality conjecture \cite{vafa}.
However there has been no trial to determine
the $ADE$ partition functions on $K3$.
This was our original motivation.
Our first trial \cite{jin3} succeeded to show the  Montonen-Olive duality,
but our $ADE$ partition functions did not satisfy the gap condition.
In \cite{jin3}
we first thought of the primary functions $G_0^{\cal G}(\tau)$
of complex coupling constant $\tau$
for $ADE$ gauge groups, 
which were derived from the denominator identity \cite{kac, mac}.
Then we obtained a set of functions generated by the modular transformations
 of the primary functions $G_0^{\cal G}(\tau)$.
Finally we prepared linear combinations of these functions 
with appropriate coefficients as the $ADE$ partition functions,
and determined the coefficients of them
by requiring the Montonen-Olive duality.
The resulting $ADE$ partition functions satisfied the Montonen-Olive duality,
but did not satisfy the gap condition.

One more difficulty of our previous proposal \cite{jin3}
is the complexity of the determination of the partition functions.
In proceeding to determine the higher rank cases of 
$ADE$ partition functions,
we find that these partition functions
are written by the deformed Hecke operator except for $D_{2N}$ and $E_6$ cases. The deformed Hecke operator is defined as follows.
We think of the theory with 't Hooft flux $v\in H^2(K3,{\bf Z}_N)$.
For meromorphic modular form $\psi(\tau)$ with weight $-12$,
the deformed Hecke operator is defined by
\beq
T_N^{\prime (v)}\psi(\tau):=\frac{1}{N^2}\sum_{
\begin{array}{c}
0\le a,b,d \in {\bf Z} \\
ad=N, b<d
\label{dho}
\end{array}
}
\delta_{dv,0}\zeta_d^{-\frac{b(\frac{v}{a})^2 }{2}}d\psi(\frac{a\tau+b}{d})
\enq
where $\delta_{v,0}$ is mod $N$ Kronecker delta
and $\zeta_d=\exp(2\pi i/d)$.
Note that this form is the same form as the Hecke operator
of order $N$ for modular form with weight $-1$ except for $\delta_{dv,0}$ and
$\zeta_d^{-\frac{b(\frac{v}{a})^2 }{2}}$ \cite{m-v,lozano,mod}.
To select the types of the block $dv=0$ mod $N$,
we introduce $\delta_{dv,0}$.
Here we point out that when $v$ satisfies 
$dv=0$ mod $ad$,
$v$ has the form $v=a{\tilde v}$(${\tilde v}\in H^2(K3,Z_d) $) 
and $(v/a)^2={\tilde v}^2=\int_{K3}{\tilde v}\wedge {\tilde v}\in 2{\bf Z} $.
 To project out the type $(v/a)^2={\tilde v}^2=2j$ mod $2d$,
we introduce $\zeta_d^{-\frac{b(\frac{v}{a})^2 }{2}}$.
These two factors are sufficient and necessary
to classify the diffeomorphic equivalent types of $v$ on $K3$ \cite{vafa-witten,jin3}.
(\ref{dho}) completely satisfies 
the full version of Montonen-Olive duality,
as we will verify later.
In the verification, we use the condition 
that $\psi(\tau)$ is meromorphic modular form with weight $-12$.
Each group has the corresponding $\psi_{\cal G}(\tau)$ as follows.
In ${\cal G}=A_r$ case, choosing $N=|\Gamma_{A_r}|=r+1$,
$\psi_{A_r}(\tau)=\frac{1}{\Delta(\tau)}$.
$\Delta(\tau)$ is 24th power of Dedekind eta function.
This result was already derived by Vafa and Witten \cite{vafa-witten,m-v,lozano}.
In ${\cal G}=D_{2N+1},E_r$ case, 
we choose $N=|\Gamma_{\cal G}|$.
$\psi_{D,E}(\tau)$ is given by the reinterpretation of our previous results \cite{jin3}
as follows.
$\psi_{\cal G}(\tau)$ is derived from the primary functions $G_0^{\cal G}(\tau)$. Concretely
$\psi_{\cal G}(\tau)$ is given by summing up all possible modular transformations of $G_0^{\cal G}(|\Gamma_{\cal G}|\tau)$.
By using  $\psi_{\cal G}(\tau)$, we define $Z_{\cal G}^v(\tau)$ by,
\beq
Z_{\cal G}^v(\tau):=T_{|\Gamma_{\cal G}|}^{\prime (v)}\psi_{\cal G}(\tau).
\label{zgv}
\enq
Note that $Z_{\cal G}^v(\tau)$ includes $G_0^{\cal G}(\tau)$
manifestly and satisfies the corresponding Montonen-Olive duality,
but does not satisfies the gap condition.

The deformed Hecke operator is a powerful tool
in the sense that we may only concentrate on 
the determination of $\psi_{\cal G}^\prime(\tau)$,
which is improved from $\psi_{\cal G}(\tau)$ 
in such a way that $T^{\prime(v=0)}_{|\Gamma_{\cal G}|}\psi_{\cal G}^\prime(\tau)$ satisfies the gap condition.
Before moving on the determination of $\psi_{\cal G}^\prime(\tau)$,
we will mention about the natural assumption 
that the total partition functions on $K3$
are holomorphic for all $ADE$ gauge groups.
This  assumption excludes the holomorphic anomaly or ${\bar \tau}$-dependence
\cite{vafa-witten,m-n,m-v,jin3} and ensures the justification of the form (\ref{dho}).
The possibility of the holomorphic anomaly is not so unnatural 
in the twisted ${\cal N}=4$ super Yang-Mills theory,
since the holomorphic anomaly comes from the failure of the reduction 
${\cal N}=4 \to {\cal N}=1$ 
by the mass perturbation proportional to the section of the canonical bundle \cite{vafa-witten}.
Typically on $\frac{1}{2}K3$, the holomorphic anomaly appears,
since $\frac{1}{2}K3$ has the locus where the section of the canonical
bundle vanishes \cite{m-v,E-S}.
On the other hand, since $K3$ has trivial canonical bundle,
we assume that the partition function on $K3$ has no holomorphic anomaly.
Now we will derive $\psi_{\cal G}^\prime(\tau)$
by improving $\psi_{\cal G}(\tau)$.
In fact the form of $\psi_{\cal G}^\prime(\tau)$
is given by the polynomial of $j(\tau)$ over $\Delta(\tau)$,
on the condition that $\psi_{\cal G}^\prime(\tau)$
has the only singularity at $i\infty$ on the upper half plane ($Im \tau >0$).
Here $j(\tau)$ is modular function.
Since the Hecke operation preserves this property,
the singularity property of $\psi_{\cal G}^\prime(\tau)$
is transferred  into that of $T_{|\Gamma_{\cal G}|}^{\prime (v=0)}\psi_{\cal G}^{\prime}(\tau)$, which is necessary for $ADE$ partition functions.
By requiring this form for $\psi_{\cal G}^\prime(\tau)$,
 we can determine $\psi_{\cal G}^\prime(\tau)$
in such a way that $T_{|\Gamma_{\cal G}|}^{\prime (v=0)}\psi_{\cal G}^{\prime}(\tau)$
satisfy the gap condition.
The resulting partition functions satisfy
the Montonen-Olive duality and the gap condition at the same time.
We expect that for all $ADE$ groups 
the above processes are valid,
but we can only derive the partition functions 
satisfying these two conditions for some groups.
Therefore we have to discuss the condition of the gapful partition functions.
In determining $\psi_{\cal G}^\prime(\tau)$
which reproduces the gapful partition functions,
we observe that the gapful partition function can exist only
when the rank of ${\cal G}$ satisfy the following property:
\beq
T_{rank({\cal G})+1}=T_{|\Gamma_{\cal G}|}(\cdots).
\label{gap}
\enq
Here $T_N$ is the true Hecke operator.
$(\cdots)$ is some polynomial of $T_M$'s which cannot be written by $T_{|\Gamma_{\cal G}|}$.
One can find 
this property of the Hecke operator in
the product of the Hecke operators \cite{mod}.
This observation is explained as follows.
 $T_{rank({\cal G})+1}\frac{1}{\Delta(\tau)}$ is always gapful
partition function, as we will show later.
In fact this gapful partition function can be obtained from $T_{|\Gamma_{\cal G}|}\psi_{\cal G}(\tau)$ in the same way as the above processes.
However these processes are not always valid 
and this partition function can be obtained from $T_{|\Gamma_{\cal G}|}\psi_{\cal G}(\tau)$ only for the cases of (\ref{gap}).
This situation is the same as that of $T_{|\Gamma_{\cal G}|}^{\prime (v=0)}\psi_{\cal G}(\tau)$.
Thus we can investigate the condition of the gapful partition functions
by using the property of the Hecke operator.

The organization of this article is the following:
In Sec.2 we review the Vafa-Witten theory and the Vafa-Witten conjecture
for the twisted ${\cal N}=4$ super Yang-Mills theory.
Next we define the deformed Hecke operator,
which satisfies the Montonen-Olive duality completely.
We reinterpret the $SU(N)$ partition functions on $K3$
by using this deformed Hecke operator.
In Sec.3 we give the universal processes of 
deriving 
our previous $D,E$ partition functions, by using the deformed Hecke operator.
In Sec.4 we improve the results in Sec.3 under the assumption that the total partition functions are holomorphic.
In Sec.5 we first introduce the gapful partition function and the product
of the Hecke operators.
By using them, we discuss the condition of the gapful partition functions.
In Sec.6. we conclude and discuss the remaining problems.

\section{Review of Vafa-Witten Theory and ${\cal N}=4~ SU(N)$ partition Functions on $K3$}
\label{sec:2}
\setcounter{equation}{0}
Vafa and Witten have first intended to test $S$-duality conjecture of ${\cal N}=4$ super Yang-Mills 
theory, by determining the exact partition functions \cite{vafa-witten}.
It is well-known that ${\cal N}=4$ super Yang-Mills theory on $4$-manifold
is exactly finite and conformally invariant.
However it is still hard to determine the exact ${\cal N}=4$ partition function itself.
Thus they thought of the twisted ${\cal N}=4$ super Yang-Mills theory,
which is topological field theory \cite{laba} and whose
partition function can be determined exactly.
In this context, they 
investigated the mathematical results of 
the $SU(N)$ partition functions on simple 
K\"ahler 4-manifolds such as $K3, C{\bf P}^2, C{\hat {\bf P}^2}$(blow-uped $C{\bf P}^2)$ and $ALE$ \cite{vafa-witten}.
On the other hand, 
$S$-duality conjecture of (twisted)${\cal N}=4$ super Yang-Mills theory is believed to be the one considered by Montonen and Olive \cite{M-O}.
This $S$-duality accompanies the exchange of gauge groups ${\cal G} \leftrightarrow{\hat {\cal G}} $.
${\hat {\cal G}}$ is the dual of ${\cal G}$.
If  ${\cal G}$ is $ADE$ gauge group, ${\hat {\cal G}}$ is given by ${\hat {\cal G}}={\cal G}/{\Gamma_{\cal G}}$. 
Here 
$\Gamma_{\cal G}$ is the center of ${\cal G}$  and $|\Gamma_{\cal G}|$ is the number of elements of $\Gamma_{\cal G}$.
To classify the theory for ${\cal G}/\Gamma_{\cal G}$,
we introduce 't Hooft fluxes $v\in H^2(X,\Gamma_{\cal G})$.
We point out that ${\cal G}/\Gamma_{\cal G}$ with $v=0$ is regarded as ${\cal G}$ itself \cite{jin3}.
\begin{table}[h]
\begin{center}
\begin{tabular}{|c|c|c|c|}
\hline
${\cal G}$& $\Gamma_{\cal G}$ & $|\Gamma_{\cal G}|$ \\
\hline
$A_{N-1}$ & ${\bf Z}_N$ &N \\
\hline
$D_{2N}$ &  ${\bf Z}_2\times{\bf Z}_2$ &4 \\
\hline
$D_{2N+1}$ &  ${\bf Z}_4$ &4 \\
\hline
$E_r, r=6,7,8$ &  ${\bf Z}_{9-r}$ &$9-r$ \\
\hline
\end{tabular}
\end{center}
\end{table} 
 
\subsection{Vafa-Witten Conjecture}

In this subsection we denote the celebrated Vafa-Witten conjecture
concerned with partition function of twisted ${\cal N}=4$ super Yang-Mills theory
on 4-manifold $X$ \cite{vafa-witten, lozano, yoshi}.
Vafa and Witten pointed out that
the partition function of twisted ${\cal N}=4$ super Yang-Mills theory
can have remarkable simple form,
if $X$ is a K\"ahler 4-manifold and vanishing theorem holds \cite{vafa-witten}.  
In this situation, its partition function has the form of 
the summation of the Euler number of the moduli space of the ASD equations.
More precisely,  for twisted ${\cal N}=4~ {\cal G}/\Gamma_{\cal G}$ theory
with 't Hooft flux $v\in H^2(X,\Gamma_{\cal G})$ on $X$,
the partition function of this theory is given by the formula, 
\begin{equation}
Z^X_v(\tau):= q^{-{\frac{(r+1)\chi(X)}{24}}}\sum_k \chi({\cal M}(v,k))q^k
\;\;\;(q:=\exp(2\pi i \tau)),\label{zxvt}
\end{equation}
where ${\cal M}(v,k) $ is the moduli space of ASD connections 
associated to ${\cal G}/\Gamma_{\cal G}$-principal bundle with 't Hooft flux $v$ and fractional instanton number $k\in \frac{1}{2|\Gamma_{\cal G}|}{\bf Z}$.
In (\ref{zxvt}), $\tau$ is the gauge coupling constant including theta angle,
$\chi(X)$ is Euler number of $X$ and $r$ is the rank of ${\cal G} $.
 The factor $q^{-{\frac{(r+1)\chi(X)}{24}}} $ is required by modular property like 
the case of $\eta(\tau) $ function.
On $K3$, $\chi(K3)=24$ and the partition function has top $q$-term $q^{-(r+1)}$.

With this result, Vafa and Witten
conjectured the behavior of the partition functions under the action of $SL(2,{\bf Z})$ on $\tau$. They started with 't Hooft's work \cite{tHooft} in mind. 
In \cite{tHooft}, the path integral with ${\bf Z}_N$-valued electric flux and 
that with magnetic flux 
are related by Fourier transform.
Vafa and Witten combined the conjecture of strong/weak duality with this
't Hooft's result. Their conjecture is summarized by the following formula:  

\begin{equation}
Z^X_v\left(-\frac{1}{\tau}\right)=\vert\Gamma_{\cal G}\vert^{-\frac{b_2(X)}{2}}
\left(
\frac{\tau}{i}
\right)^{-\frac{\chi(X)}{2}}
\sum_u\zeta_{|\Gamma_{\cal G}|}^{u\cdot v}Z_u^X(\tau),
\label{m-o}
\end{equation}
where $\zeta_{|\Gamma_{\cal G}|}=\exp(2\pi i/|\Gamma_{\cal G}|)$ \cite{vafa-witten}.

Hereafter we will concentrate on $X=K3$ case.
We discuss the gap condition on $K3$.
In $v=0$ case
${\cal M}(0,k):={\cal M}_k^{\cal G}  $ is  the moduli space of irreducible 
ASD connections associated to ${\cal G}$-principal bundle with instanton number $k$.
Its dimension $\dim {\cal M}_k^{\cal G}$ is given by
Atiyah-Hitchin-Singer dimension formula \cite{AHS}:
 \beq
\dim{\cal M}^{\cal G}_k=4h({\cal G})(k-r)-4r,
\label{ahs}
\enq 
where $h({\cal G})$ is the dual Coxeter number, $\dim{\cal G}$ is the dimension
of ${\cal G}$ and $k$ is the instanton number. 
$ \dim{\cal M}^{\cal G}_k\ge 0$ means $k\ge r+1$.
That is, the moduli space of irreducible ASD connections with 
$ADE$ gauge group ${\cal G}$ on $K3$ exist only for $k\ge r+1$ except for $k=0$ trivial case.
This condition restricts the form of the partition function. 
That is, the ${\cal G}$ partition function cannot have $q^{k-(r+1)}$ terms for $1\le k \le r$.
This condition is called gap condition \cite{vafa-witten, m-n, m-v, E-S, jin3}.

\subsection{Deformed Hecke Operator and $SU(N)$ Partition Functions on $K3$}

In this subsection, we define the deformed Hecke Operator, 
which reproduces Montonen-Olive duality for  $SU(N)$ partition functions on $K3$.
In \cite{vafa-witten}, Vafa and Witten classified the types of 
$SO(3)=SU(2)/{\bf Z}_2$ 
theory with 't Hooft flux $v\in H^2(K3,{\bf Z}_2)$ on $K3$. 
There are three types:
$v^2= 0$ mod 4 and $v\ne 0$ (even),$v^2= 2$ mod 4 (odd) and $v=0$(trivial).
Here it is convenient to introduce the intersection matrix on $K3$ 
by $H^{\oplus 11}$ instead of $H^{\oplus 3}\oplus (-E_8)^{\oplus 2}$ \cite{Fukaya}.
Here 
\beq
H=\left(\begin{array}{cc}
0&1\\
1&0
\end{array}\right).
\enq
This replacement is sufficient to think of the partition functions 
with $v\in H^2(K3,{\bf Z}_2)$ \cite{vafa-witten}.
The generalization to $SU(N)/{\bf Z}_N$  theory with $v\in H^2(K3,{\bf Z}_N)$ is straightforward,
but there is a complicated structure for general $N$. 
To explain this structure, we introduce the block and the type \cite{jin3}.
For general $N$ with divisors $\{ d_i \}$($a_i=N/d_i$),
the block $B_{d_i}$ is defined by
\beq
B_{d_i}:=\{v| d_iv=0 \mbox{ and } {}^\forall d_jv\ne 0 \mbox{ mod } N, d_i|d_j 
\mbox{ and } d_i\ne d_j  \}.
\enq
For $v\in B_{d_i}$, we can introduce $v=a_i{\tilde v}({\tilde v}\in H^2(K3,{\bf Z}_{d_i}))$. 
The block $B_{d_i}$ has $d_i$ types:
\beq
[v_j]:=\{v| (v/a_i)^2 ={\tilde v}^2=\int_{K3}{\tilde v}\wedge {\tilde v}=2j \mbox{ mod } 2d_i \}.
\enq
We call these types and these structure as orbits and orbit structures
respectively.
The numbers of the orbit are counted concretely in \cite{jin3}.
For the theory with $v\in H^2(K3,{\bf Z}_N)$, we define the deformed Hecke operator by
\beq
T_N^{\prime (v)}\psi(\tau):=\frac{1}{N^2}\sum_{
\begin{array}{c}
0\le a,b,d \in {\bf Z} \\
ad=N,b<d
\label{dho4}
\end{array}
}
\delta_{dv,0}\zeta_d^{-\frac{b(\frac{v}{a})^2 }{2}}d\psi(\frac{a\tau+b}{d}),
\enq
where $\delta_{v,0}$ is mod $N$ Kronecker delta
and $\zeta_d=\exp(2\pi i/d)$.
$\psi(\tau)$ is meromorphic modular form with weight $-12$.
$v=0$ reduces to $T_N^{\prime(v=0)}$, which has the same form as 
the Hecke operator of order $N$ for modular weight $-1$ \cite{m-v,lozano}. 
We introduce $\delta_{v,0}$ and $\zeta_d$ factors,
in such a way that this operator can represent the  partition function of 
all blocks and types \cite{jin3}.
$\delta_{dv,0}$ selects the block $dv=0$ mod $2N$,
and $\zeta_d^{-\frac{b(\frac{v}{a})^2 }{2}}$ projects the type $(v/a)^2=2j$ mod $2d$.
For this definition, we are inspired by the results in \cite{vafa-witten,lozano,iqbal,jin3}.

The deformed Hecke operator has the following modular property:
\beq
T_N^{\prime (v)}\psi\left(-\frac{1}{\tau}\right)=N^{-11}\tau^{-12}\sum_u\zeta_N^{u\cdot v}
T_N^{\prime (u)}\psi({\tau}).
\label{mpdho}
\enq
This is nothing but the Montonen-Olive duality (\ref{m-o}).
Now we will verify (\ref{mpdho}).
L.H.S. of (\ref{mpdho}) is given by
\beq
T_N^{\prime (v)}\psi\left( -\frac{1}{\tau}\right)
=N^{-11}\tau^{-12}
\frac{1}{N^2}\sum_{
\begin{array}{c}
0\le a,b,d \in {\bf Z} \\
ad=N,b<d
\label{mp1}
\end{array}
}
\delta_{dv,0}\zeta_d^{-\frac{b(\frac{v}{a})^2 }{2}}
\left(\frac{N}{p} \right)^{12}a^{11}
\psi(\frac{p\tau+ab^\prime}{a{\tilde d}}),
\enq
where
$p=\mbox{gcd}(b,d),{\tilde d}=d/p,{\tilde b}=b/p,b^\prime{\tilde d}=-1
\mbox{ mod } {\tilde d}$.
If $b=0$, then $p=d$ and $b^\prime=0={\tilde b}$ \cite{lozano}.
To derive R.H.S. of (\ref{mpdho}), we give the equivalent form to (\ref{dho4}),
\beq
T_N^{\prime (v)}\psi(\tau):=\frac{1}{N^2}\sum_{
\begin{array}{c}
0\le a,p,b^\prime,{\tilde d} \in {\bf Z} \\
ap{\tilde d}=N,b^\prime<{\tilde d}
\label{dho2}
\end{array}
}
\delta_{a{\tilde d}v,0}\zeta_{a{\tilde d}}^{-\frac{ab^\prime(\frac{v}{p})^2 }{2}}a{\tilde d}\psi(\frac{p\tau+ab^\prime}{a{\tilde d}}).
\enq
By using this, R.H.S. of (\ref{mpdho}) is given by
\beqy
\sum_{u}\zeta_N^{u\cdot v}T_N^{\prime (u)}\psi(\tau)
&=&\sum_u \zeta_N^{u\cdot v}
\frac{1}{N^2}\sum_{
\begin{array}{c}
0\le a,p,b^\prime,{\tilde d} \in {\bf Z} \\
ap{\tilde d}=N,b^\prime<{\tilde d}
\label{dho1}
\end{array}
}
\delta_{a{\tilde d}u,0}\zeta_{a{\tilde d}}^{-\frac{ab^\prime(\frac{u}{p})^2 }{2}}a{\tilde d}\psi(\frac{p\tau+ab^\prime}{a{\tilde d}})
\no\\
&=&
\frac{1}{N^2}\sum_{
\begin{array}{c}
0\le a,p,b^\prime,{\tilde d} \in {\bf Z} \\
ap{\tilde d}=N,b^\prime<{\tilde d}
\label{mp2}
\end{array}
}
\delta_{a{\tilde d}v,0}\zeta_{a{\tilde d}}^{-\frac{a{\tilde b}(\frac{v}{p})^2 }{2}}
(a^2{\tilde d})^{11}a{\tilde d}\psi(\frac{p\tau+ab^\prime}{a{\tilde d}}).
\enqy
To obtain the last equality, we use the identity,
\beq
\sum_u \zeta_N^{u\cdot v}\delta_{a{\tilde d}u,0}\zeta_{a{\tilde d}}^{-\frac{ab^\prime}{2}(\frac{u}{p})^2}
=\delta_{a{\tilde d}v,0}\zeta_{a{\tilde d}}^{-\frac{a{\tilde b}}{2}(\frac{v}{p})^2}(a^2{\tilde d})^{11}.
\enq
This identity is verified in Appendix A.
(\ref{mp1}) and (\ref{mp2}) show the equivalence between (\ref{mp1}) and (\ref{mp2}).
We point out that the modular property (\ref{mpdho}) is similar to that of level $N$ affine Lie algebra \cite{kac}. For $SU(N)$, we have already pointed out that the $SU(N)$ partition functions are related to the level $N$ affine $SU(N)$ algebra in \cite{jin2}. 

By using the deformed Hecke operator,
we can define the $SU(N)/{\bf Z}_N$ partition functions with $v\in H^2(K3,{\bf Z}_N)$ on $K3$ as follows.
We request (\ref{zxvt}) for the $SU(N)$ partition functions
written by the deformed Hecke operator.
(\ref{zxvt}) means that the $SU(N)$ partition function has top $q$-term $1/q^N$. Since $T_N^{\prime (v=0)}\frac{1}{\Delta} \sim 1/q^N$,we can conclude $\psi_{A_{N-1}}(\tau)=\frac{1}{\Delta(\tau)}$.
Thus we can obtain the $SU(N)/{\bf Z}_N$ partition functions with $v\in H^2(K3,{\bf Z}_N)$:
\beq
Z_{A_{N-1}}^v(\tau)=T^{\prime (v)}_N\frac{1}{\Delta(\tau)}.
\enq
We have to check the gap condition for $SU(N)/{\bf Z}_N$ partition functions with $v=0$.
As we will show in Sec.5, $T^{\prime (v=0)}_N\frac{1}{\Delta(\tau)}$ has a gap.
For ${\cal G}=A_{N-1}$ case, we can obtain $SU(N)$ partition functions,
which satisfy the Montonen-Olive duality and the gap condition at the same time,
when we only request the condition of top $q$-term for the partition function written by the deformed Hecke operator. 

\section{$D,E$ Partition Functions on $K3$ without Gap}
\label{sec:2}
\setcounter{equation}{0}
In this section, we rederive the previous results \cite{jin3} 
by using the deformed Hecke operator.
For this purpose, we first introduce the primary function $G_0^{\cal G}(\tau)$
 for $ADE$ gauge groups.
Next we reproduce the previous results by using these primary functions.

\subsection{Denominator Identity and Stringy Picture of ${\cal N}=4~ADE$ Theory on $K3$}
In this subsection, we introduce the primary functions $G_0^{\cal G}(\tau)$
for $ADE$ gauge groups, which are necessary 
not only for the previous derivations \cite{jin3}
but also for the present derivations.
We will use the primary functions as a piece of 
$ADE$ partition functions on $K3$.
We begin with the following stringy picture.\\
 \begin{tabular}{ccc}
\\
IIA on $K3\times T^2 \times ALE_{\small ADE}$ &$\leftrightarrow$& Hetero on $T^4\times T^2 \times ALE_{ADE}$\\
$\downarrow$ & &$\downarrow$\\
${\cal N}=4 ~ADE$ on $K3$ &$\leftrightarrow$& (${\cal N}=4~ U(1)$ on $ALE_{ADE})^{\otimes 24}$
\\
\\
\end{tabular} 
\\First line is IIA/Hetero duality \cite{hj}. In second line of IIA side, we compactify
$T^2\times ALE_{ADE}$ and obtain ${\cal N}=4 ~ADE$ on $K3$.
$ADE$ gauge group is determined by the type of the $ALE$ space,
which is classified by the $ADE$ sub-group of $SU(2)$ \cite{vafa}.
In second line of Hetero side, we compactify $T^4\times T^2$
and obtain $({\cal N}=4~U(1)$ on $ALE_{ADE})^{\otimes 24}$ \cite{jin3}.
The origin of ${\otimes 24}$ is explained by 
the compactification of $K3\times T^2$
in IIA side.
Thus each side of 
the second line is equivalent and their partition functions 
in both side are the same.
On the other hand, the partition function of ${\cal N}=4~U(1)$ theory on $ALE_{ADE}$
was already obtained by Nakajima \cite{naka}.
In\cite{jin3}, we introduced the $ADE$ blow-up formula $\theta^2_{\cal G}(\tau)/\eta(\tau)^{r+1}$,
which is very similar to Nakajima's results.
This blow-up formula was derived from the generalization of our previous
results for $SU(N)$ \cite{jin, jin2}.
Explicit $\theta^2_{\cal G}(\tau)/\eta(\tau)^{r+1}$ is given as follows:
\begin{table}[h]
\begin{center}
\begin{tabular}{|c|c|c|}
\hline
${\cal G}$& $\theta^2_{\cal G}(\tau)/\eta(\tau)^{r+1}$\\
\hline
$A_{r}$ & {\Large$\frac{1}{\eta(\frac{\tau}{r+1})}$} \\
\hline
$D_{r}$ &  {\Large$\frac{\eta(\frac{\tau}{r-1})}{\eta(\frac{\tau}{2})\eta(\frac{\tau}{2r-2})}$}\\
\hline
$E_6$ &  {\Large$\frac{\eta(\frac{\tau}{4})\eta(\frac{\tau}{6})}{\eta(\frac{\tau}{2})\eta(\frac{\tau}{3})\eta(\frac{\tau}{12})}$}\\
\hline
$E_7$ &  {\Large$\frac{\eta(\frac{\tau}{6})\eta(\frac{\tau}{9})}{\eta(\frac{\tau}{2})\eta(\frac{\tau}{3})\eta(\frac{\tau}{18})}$}\\
\hline
$E_8$ &  {\Large$\frac{\eta(\frac{\tau}{6})\eta(\frac{\tau}{10})\eta(\frac{\tau}{15})}{\eta(\frac{\tau}{2})\eta(\frac{\tau}{3})\eta(\frac{\tau}{5})\eta(\frac{\tau}{30})}$}\\
\hline
\end{tabular}
\end{center}
\end{table} \\
We remark that all $\theta^2_{\cal G}(\tau)/\eta(\tau)^{r+1} $ are eta-product.
This fact comes from the definition of $\theta^2_{\cal G}(\tau)$.
In \cite{jin3}, we defined $\theta^2_{\cal G}(\tau)$
by using the affine Lie algebra \cite{kac, mac}.
By combining the relations of stringy picture with $\theta^2_{\cal G}(\tau)/\eta(\tau)^{r+1} $,
we define the primary functions for $ADE$ gauge groups on $K3$: 
\beq
G_0^{\cal G}(\tau):=\left( \frac{\theta^2_{\cal G}(\tau)}{\eta^{r+1}(\tau)}\right)^{24}.
\enq
For $SU(N)$ theory, this form is consistent with the fact that the moduli space of rank $N$ semi-stable sheaves is described by Hilbert scheme of $K3$ \cite{mukai,nak,yoshihecke}.

\subsection{$D_{2N+1}$ Partition Functions on $K3$ without Gap}
We will first sketch the processes of the determinations 
of the partition functions in our previous work \cite{jin3}.
In \cite{jin3}, we introduced a set of functions generated
by the modular transformations of $G_0^{\cal G}(\tau)$
(in some cases we also included functions generated
by the modular transformations $\tau\to \tau+1/m$).
We defined the $D,E$ partition functions 
by the linear combinations of these functions with appropriate coefficients.
Finally we determined the coefficients in the partition functions,
so that these partition functions satisfy the corresponding Montonen-Olive
duality (\ref{m-o}).
 
In deriving the higher rank cases of $D,E$,
we observe that the $D,E$ partition functions are also described 
by the deformed Hecke operator for some groups.
In $D_N$ case, $D_{2N+1}$ partition  functions
 can be described by the deformed Hecke operator, 
 while $D_{2N}$ partition functions cannot.
Since the Montonen-Olive duality for $D_{2N}$ is complicated \cite{jin3},
we do not treat these cases in this article.

We find the general processes to define the $D,E$ partition functions.
They are equivalent to the previous processes \cite{jin3},
but more general than before.
In $D_{2N+1}$ case, from the primary functions
we define $\phi^{D_{2N+1}}(\tau)$ by   
 \beq
\phi^{D_{2N+1}}(\tau):=G_0^{D_{2N+1}}(4\tau).
\label{phi}
\enq
$4\tau$ in (\ref{phi}) comes from $|\Gamma_{D_{2N+1}}|=4$. 
Then we define the meromorphic modular forms $\psi_{D_{2N+1}}(\tau)$
with weight $-12$ by
\beq
\psi_{D_{2N+1}}(\tau):=\frac{1}{2}\sum_{\mbox {\tiny all possible modular transformations }}
\phi^{D_{2N+1}}(\tau),
\enq
where we sum up  all possible modular transformations
of $\phi^{D_{2N+1}}(\tau)$ with $\tau^{-12}$ factors etc. neglected.
Overall factor $1/2$ is introduced to compare the partition functions
with those in Sec.5.
By using $\psi_{D_{2N+1}}(\tau)$, we define the $D_{2N+1}$ partition functions with
$v\in H^2(K3,{\bf Z}_4)$ by
\beq
Z_{D_{2N+1}}^{v}(\tau):=T_{4}^{\prime(v)} \psi_{D_{2N+1}}(\tau).
\enq
Here we remark that $Z_{D_{2N+1}}^{v}(\tau) $ contains $G_0^{D_{2N+1}}(\tau)$ automatically. We give some examples as follows,
\beq
\psi_{D_3}(\tau)=\frac{1}{\Delta},
\enq

\beq
\psi_{D_5}(\tau)=\frac{24}{\Delta},
\enq

\beq
\psi_{D_7}(\tau)=\frac{1}{\Delta}(j+6),
\enq

\beq
\psi_{D_9}(\tau)=\frac{1}{\Delta}(24j-9984),
\enq

\beq
\psi_{D_{11}}(\tau)=\frac{1}{\Delta}(j^2-1188j+9),
\enq

\beq
\psi_{D_{13}}(\tau)=\frac{1}{\Delta}(24j^2-33088j+2304192),
\enq

\beq
\psi_{D_{15}}(\tau)=\frac{1}{\Delta}(j^3-1932j^2+641682j+12),
\enq

\beq
\psi_{D_{17}}(\tau)=\frac{1}{\Delta}(24j^3-50944j^2+21879936j-531714048),
\enq

\beq
\psi_{D_{19}}(\tau)=\frac{1}{\Delta}(j^4-2676j^3+1882206j^2-266891000j+18),
\enq

\beq
\psi_{D_{21}}(\tau)=\frac{1}{\Delta}(24j^4-68800j^3+55057056j^2-10821624576j+122700718368).
\enq
Here $j(\tau)=E_4^3(\tau)/\Delta(\tau)$.\\
$D_{2N+1}$ partition functions satisfy the Montonen-Olive duality 
(\ref{m-o}) by construction.
What about other conditions (\ref{zxvt}),(\ref{ahs}) ?
$D_{2N+1}$ partition functions $Z_{D_{2N+1}}^{v=0}(\tau) $ do not satisfy
the condition (\ref{ahs}).
We will improve this point in the next section.
$D_{4N+3}$ partition functions satisfy the condition (\ref{zxvt}),
while $D_{4N+1}$ partition functions do not satisfy this.
In $D_{4N+1}$, the partition functions of the previous processes \cite{jin3},
which are derived from $G_0^{D_{4N+1}}(\tau)$ directly,
contains $q^{-(4N+2)}$ terms formally.
However these terms are mutually cancelled.
Thus  $D_{4N+1}$ partition functions 
starts from $q^{-4N}$ instead of $q^{-(4N+2)}$.
This is explained as follows.
Since all $D_{2N+1}$ partition functions
are described by the deformed Hecke operator $T_4^{\prime(v=0)}$,
top $q$-term must have the form $q^{-4M}$,
which is invariant under the translation $\tau\to \tau+1/4$.
As a result
we find that there are two classes $D_{4N+1}$ and $D_{4N+3}$
in $D_{2N+1}$.

\subsection{$E_{7,8}$ Partition functions on $K3$ without Gap}
The processes of $E_{7,8}$ cases are the same as those of $D_{2N+1}$.
\paragraph{ A) $E_7$}~\\
By replacing $4\tau$ in (\ref{phi}) by $2\tau$,$\phi^{E_7}(\tau)$ of $E_7$ is given by 
\beq
\phi^{E_7}(\tau):=G_0^{E_7}(2\tau).
\enq
By using this, we obtain
\beq
\psi_{E_7}(\tau):=\frac{1}{2}\sum_{\mbox {\tiny all possible modular transformations }}
\phi^{E_7}(\tau).
\enq
Finally we can define the $E_7$ partition functions with $v\in H^2(K3,{\bf Z}_2)$ by 
\beq
Z_{E_7}^v(\tau):=T_{2}^{\prime(v)} \psi_{E_7}(\tau),
\enq
\beq
\psi_{E_7}(\tau)=\frac{1}{\Delta}(j^3-\frac{4465}{2}j^2+1070688j-36935982).
\enq
\paragraph{B) $E_8$}~\\
$E_8$ case are also treated in the same way.
\beq
\phi^{E_8}(\tau):=G_0^{E_8}(\tau),
\enq
\beqy
Z_{E_8}(\tau)&=&\psi_{E_8}(\tau)=\sum_{\mbox {\tiny all possible modular transformations }}
\phi^{E_8}(\tau)
\no
\\
&=&\frac{1}{\Delta}(
j^8-5976j^7
+14049204j^6-16450492296j^5
\no\\&&
+10006744823442j^4
-2995100782701144j^3
+373127947258066100j^2
\no\\&&
-12808385327808647208j+26763599994092029512).
\enqy
$E_7$ and $E_8$ both satisfy the conditions ($\ref{zxvt}$) and (\ref{m-o}),
but do not satisfy the gap condition (\ref{ahs}).

We will give a comment on $E_6$ case.
Formally $E_6$ case can be treated in the same way as $D_{2N+1}$ and $E_{7,8}$.
However $\psi_{E_6}(\tau)$ starts from $q^{-3}$ 
and $Z_{E_{6}}^{v=0}(\tau)$ starts from $q^{-9}$,which
fails to satisfy the condition (\ref{zxvt}).
Thus we do not consider $E_6$ case in this article.

\section{$D,E$ Partition Functions on $K3$ with Gap}
\label{sec:4}
\setcounter{equation}{0}
In the previous section, we derived the $D,E$ partition functions
described by the deformed Hecke operator,
\beq
Z_{{\cal G}}^v(\tau):=T_{|\Gamma_{\cal G}|}^{\prime(v)} \psi_{\cal G}(\tau).
\enq
However they do not satisfy the gap condition. We want to improve these functions in such a way that they satisfy the gap condition.
Concretely we want to subtract some functions from these functions,
in such a way that the total partition functions satisfy the gap condition and
the Montonen-Olive duality at the same time.
What kind of functions can we subtract ?
We assume that the total functions are holomorphic.
Thus the functions to be subtracted also must be holomorphic.
If the partition function is not holomorphic,
we would not determine it only with the deformed Hecke operator.
When the twisted $N=4$ partition function has holomorphic anomaly,
it has ${\bar \tau}$-dependence and is not holomorphic \cite{vafa-witten,m-n,m-v,E-S}.
The twisted $N=4$ partition function on K\"ahler 4-manifold
can be reduced to the twisted $N=1$ partition function,
which is easier to determine.
This reduction is done by the holomorphic deformation
using the mass perturbation proportional to the section of the canonical bundle \cite{vafa-witten}.
However there are cases when this section vanishes.
In these cases, one has to consider the contribution from the holomorphic anomaly \cite{vafa-witten,m-v}.   
$\frac{1}{2}K3$ or $C{\bf P}^2$ are these cases.
On the other hand, there is no locus where the section of the canonical bundle vanishes, since $K3$ has the trivial canonical bundle.
Thus the partition function on $K3$ has no holomorphic anomaly and is holomorphic.
Once we assume that the functions to be subtracted are holomorphic,
it is easy to assume that the form of the functions to be subtracted
satisfies the Montonen-Olive duality.
The deformed Hecke operator makes meromorphic modular forms with weight $-12$
satisfy the Montonen-Olive duality automatically.
We only have to determine $\psi^\prime_{\cal G}(\tau)$ 
which satisfies two conditions, by subtracting some meromorphic modular forms with weight $-12$.
To determine  $\psi^\prime_{\cal G}(\tau)$, we restrict the form of it 
by requiring the singularity property of the partition functions as follows.
$\psi_{\cal G}(\tau)$ has the only singularity at $i \infty$ on upper half-plane($Im\tau>0$).
This property is kept under the operation of the deformed Hecke operators.
That is, $T_N^{\prime (v=0)}\psi_{\cal G}(\tau)$ also has the singularity 
at $i \infty$ on upper half-plane and does not have the other singularities.
We also request the same property for $\psi^\prime_{\cal G}(\tau)$
even after the subtraction.
$i\infty$ on upper half-plane($Im\tau>0$) is corresponding to the weak coupling limit $Im \tau=4\pi/g^2\to \infty$
after we deform the twisted ${\cal N}=4$ Yang-Mills theory to ${\cal N}=2$ or ${\cal N}=1$
theory by the appropriate mass perturbation.
It is well-known that ${\cal N}=2$ or ${\cal N}=1$
theory is asymptotic free.
We optimistically assume that the reduction ${\cal N}=4\to{\cal N}=1$ would not produce the other singularities except for $i\infty$.
If the meromorphic modular form with weight $-12$ has the only singularity at $i\infty$
on upper half-plane,
the form of it is only the polynomial of $j(\tau)$ over $\Delta(\tau)$.
Note that negative $j(\tau)$ powers can produce the other singularities.
Finally we define the gapful partition functions from the original one by
\beq
Z_{{\cal G}}^{\prime v}(\tau):=T_{|\Gamma_{\cal G}|}^{\prime (v)}\psi^\prime_{\cal G}(\tau),
\label{zpg}
\enq
\beqy
\psi^\prime_{\cal G}(\tau)&:=&\psi_{\cal G}(\tau)+\frac{1}{\Delta}\sum_{k=0}^{[\frac{r+1}{|\Gamma_{\cal G}|}]-2}a_k^\prime j^k
\no\\
&=&\frac{1}{\Delta}\sum_{k=0}^{[\frac{r+1}{|\Gamma_{\cal G}|}]-1}a_kj^k
\label{zp}
\enqy
where we determine $a_k^\prime (a_k)$ 
in such a way that the partition functions satisfy the gap condition. \\
Seeing the form (\ref{zp}),
one would wonder the role of $\psi_{\cal G}(\tau)$ in  $\psi^\prime_{\cal G}(\tau)$.
That is, $\psi_{\cal G}(\tau)$ would not be necessary to determine $\psi^\prime_{\cal G}(\tau)$ which reproduces the gapful partition functions.
Indeed one can determine $\psi^\prime_{\cal G}(\tau)$ directly from
the polynomial of $j(\tau) $ over $\Delta(\tau)$.
However we believe that $\psi^\prime_{\cal G}(\tau)$ and $\psi^\prime_{\cal G}(\tau)-\psi_{\cal G}(\tau)$ must have the different roles in $Z_{\cal G}^{\prime (v)}(\tau)$,
since $\psi_{\cal G}(\tau)$ itself is interpreted with stringy picture in Sec.3. Interpretation of $\psi^\prime_{\cal G}(\tau)-\psi_{\cal G}(\tau)$
is the remaining problem.

\subsection{$D_{2N+1}$ Partition Functions on $K3$ with Gap }
Following the above processes, we define the $D_{2N+1}$ partition functions
with gap,
\beq
Z_{D_{2N+1}}^{\prime v}(\tau):=T_{4}^{\prime (v)}\psi^\prime_{D_{2N+1}}(\tau),
\enq
where $\psi^\prime_{D_{2N+1}}(\tau)$ are determined so that $Z_{D_{2N+1}}^{\prime v=0}(\tau)$ have gap.
Explicit $\psi^\prime_{D_{2N+1}}(\tau)$'s are given by,

\beq
\psi_{D_3}^\prime(\tau)=\frac{1}{\Delta},
\enq

\beq
\psi_{D_5}^\prime(\tau)=24\psi_{D_3}^\prime(\tau),
\enq

\beq
\psi_{D_{11}}^\prime(\tau)=\frac{1}{\Delta}(j^2-1512j+177876),
\enq

\beq
\psi_{D_{13}}^\prime(\tau)=24\psi_{D_{11}}^\prime(\tau),
\enq

\beq
\psi_{D_{19}}^\prime(\tau)=\frac{1}{\Delta}(j^4-3000j^3+2587500j^2-587500000j+9433593750),
\enq

\beq
\psi_{D_{21}}^\prime(\tau)=24\psi_{D_{19}}^\prime(\tau),
\enq

\beqy
\psi_{D_{27}}^\prime(\tau)&=&\frac{1}{\Delta}(j^6-4488j^5+7211268j^4-4959991520j^3
\no\\
&&
+1353521229474j^2-102751222086864j+502355677519592),
\enqy

\beq
\psi_{D_{29}}^\prime(\tau)=24\psi_{D_{27}}^\prime(\tau),
\enq

\beqy
\psi_{D_{31}}^\prime(\tau)&=&\frac{1}{\Delta}(j^7-5232j^6+10353456j^5-9609534976j^4+4197252771840j^3
\no\\
&&
-751984016228352j^2+37376799281254396j
\no\\&&
-1159259000376142848),
\enqy

\beq
\psi_{D_{33}}^\prime(\tau)=24\psi_{D_{31}}^\prime(\tau).
\enq
Note that we only listed up a part of $D_{2N+1}$'s.
As explained in the next section,
we can only make a part of $D_{2N+1}$ partition functions gapful.
We also note that $\psi_{D_{4N+5}}^\prime(\tau)=24\psi_{D_{4N+3}}^\prime(\tau)$. $\psi_{D_{4N+5}}^\prime(\tau)$ and $\psi_{D_{4N+3}}^\prime(\tau)$
belong to the same universality class that has the same top $q$ power.
\subsection{$E_{7,8}$ Partition Functions on $K3$ with Gap }
$E_7$ and $E_8$ cases follow the same processes,
\beq
Z_{E_7}^{\prime v}(\tau):=T_2^{\prime (v)}\psi_{E_7}^\prime(\tau),
\enq

\beq
\psi_{E_7}^\prime(\tau)=\frac{1}{\Delta}(j^3-2256j^2+1105920j-40890372),
\enq

\beqy
Z_{E_8}^\prime(\tau)=\psi_{E_8}^\prime(\tau)&=&\frac{1}{\Delta}(
j^8-5976j^7+14049180j^6-16450384608j^5
\no\\&&
+10006571842542j^4-2994981852876048j^3+373095512069721768j^2
\no\\&&
-12805925359104567360j+26751592671298309053)
\\&=& 9^{13}T_9\frac{1}{\Delta}.\label{e8}
\enqy
Here $T_9$ is the true Hecke operator, as we will introduce later.
The last equality (\ref{e8}) seems mysterious.
However as we will explain in the next section,
$9^{13}T_9\frac{1}{\Delta}$ and $Z_{E_8}^\prime(\tau)$ belong to the same universality class.
\section{Product of Hecke Operator and Condition to Gapful Partition Functions}
\label{sec:5}
\setcounter{equation}{0}
In the previous section, we derived the $D,E$ partition functions with gap.
However in some $D,E$ cases we could not make the $D,E$ partition functions gapful.
In this section, we discuss the condition of the gapful partition functions
by using the properties of Hecke operators.
In this section, we only think of $D_{4N+3}$ in $D_{2N+1}$, 
since we have already seen $\psi^\prime_{D_{4N+5}}(\tau)=24\psi^\prime_{D_{4N+3}}(\tau)$.

To discuss the condition of the gapful partition functions,
we introduce the following partition functions
by replacing $T^{\prime (v)}_{|\Gamma_{\cal G}|}$ by $T_{|\Gamma_{\cal G}|}$
and $\psi^\prime_{\cal G}$ by ${\tilde \psi}_{\cal G}$ in (\ref{zpg}),
\beq
{\tilde Z}_{{\cal G}}(\tau):=T_{|\Gamma_{\cal G}|}{\tilde \psi}_{\cal G}(\tau),
\label{zt}
\enq
\beqy
{\tilde \psi}_{\cal G}(\tau)&:=&\psi_{\cal G}(\tau)+\frac{1}{\Delta}\sum_{k=0}^{[\frac{r+1}{|\Gamma_{\cal G}|}]-2}{\tilde a}_k j^k
\no\\
&=&\frac{1}{\Delta}\sum_{k=0}^{[\frac{r+1}{|\Gamma_{\cal G}|}]-1}b_kj^k,
\label{pt}
\enqy
where ${\tilde \psi}_{\cal G}(\tau) $ is also determined
in such a way that ${\tilde Z}_{\cal G}(\tau) $ are gapful.
$T_{N}$ is the true Hecke operator \cite{mod} defined by,
\beq
T_N\psi(\tau):=\frac{1}{N^{13}}\sum_{
\begin{array}{c}
0\le a,b,d \in {\bf Z} \\
ad=N,b<d
\label{tho}
\end{array}
}
d^{12}\psi(\frac{a\tau+b}{d}).
\enq
$T_N^{\prime (v=0)}$ and $T_N$ are only different in the power of $d$
except for overall factor.
Starting from the same $\psi_{\cal G}(\tau)$,
whether $T^{\prime (v=0)}_{|\Gamma_{\cal G}|}\psi_{\cal G}(\tau)$ 
or $T_{|\Gamma_{\cal G}|}\psi_{\cal G}(\tau)$ can be made gapful or not,
only depends on ${\cal G}$  in our calculations.
So, to discuss the condition of the gapful partition functions,
we use ${\tilde Z}_{\cal G}(\tau)$ instead of $Z_{\cal G}^{\prime v=0}(\tau)$ itself.
When can we obtain  ${\tilde Z}_{\cal G}(\tau)$ ?
We discuss this point by using the product of Hecke operators.
Before moving on $D,E$ cases, 
we briefly show why $A_r$ partition functions are always gapful.
That is, we show that we can always define ${\tilde Z}_{A_r}(\tau)$. 
For this purpose, we give an important observation,
\beq
{\tilde Z}_{{\cal G}}(\tau)=T_{(rank({\cal G})+1)}\frac{1}{\Delta(\tau)}.
\label{ob}
\enq
When ${\cal G}=A$,
this observation shows that ${\tilde \psi}_{A_r}(\tau)=\psi_{A_r}(\tau)=\frac{1}{\Delta(\tau)}$ since $rank(A_r)+1=|\Gamma_{A_r}|$.
This implies that we need not subtract functions from the original one.
R.H.S. includes the summation
\beqy
\sum_{b=0}^{d-1}\frac{1}{\Delta(\frac{a\tau+b}{d})}
&=&q^{-\frac{a}{d}}(1+\zeta_d^{-1}+\zeta_d^{-2}+\cdots + \zeta_d^{-(d-1)})+{\cal O}(q^0)+\cdots,
\no\\
&=&
\left\{
\begin{array}{l}
{\cal O}(q^0)+\cdots,(d>1)\\
q^{-a}+{\cal O}(q^0)+\cdots,(d=1)
\end{array}
\right. .
\enqy
Thus ${\tilde Z}_{A_r}(\tau)$ can always be defined. The observation (\ref{ob})
is also valid for $D,E$ cases.
But $T_{(rank({\cal G})+1)}$ is not $T_{|\Gamma_{\cal G}|}$
when ${\cal G}=D,E$ (${\cal G}\ne D_3$).
To identify (\ref{ob}) with (\ref{zt}), we search the cases when
$T_{(rank({\cal G})+1)}$ has the factor $T_{|\Gamma_{\cal G}|}$.
For this purpose, we cite the following theorem \cite{mod},
\begin{thm}
(Hecke)\\
For $\psi(\tau)$ with modular weight $-12 $,
the Hecke operators satisfy
\begin{itemize}
\item[(a)] For a primitive $p^r$ with $r\ge 1$,
\beq
T_{p^r}T_p \psi(\tau)=T_{p^{r+1}}\psi(\tau)+p^{-13}T_{p^{r-1}}\psi(\tau).
\enq
Hence $T_{p^r}$ is a polynomial in $T_p$ with integer coefficients.
\item[(b)]
\beq
T_m T_n \psi(\tau)=T_{mn}\psi(\tau)
\enq
if $m$ and $n$ are relatively prime.
\item[(c)] The algebra generated by the $T_n$ for $n=1,2,3,\ldots $
is generated by the $T_p$ with $p$ prime and is commutative.
\end{itemize}
\end{thm}

Hereafter we list up the cases when $T_{rank({\cal G})+1}$ has the factor $T_{|\Gamma_{\cal G}|}$ for $D_{4N+3}$ and $E_{7,8}$ cases.
\paragraph{a) $D_{4N+3}$}~ \\
By using the theorem we find
\beq
T_{2^{2+3M}(2N^\prime+1)}\frac{1}{\Delta}= T_4(\mbox{polynomial of }T_2)T_{2N^\prime+1}\frac{1}{\Delta}.
\enq
L.H.S. is (\ref{ob}) and R.H.S. is (\ref{zt}).
Thus we can read 
\beq
{\tilde \psi}_{D_{2^{2+3M}(2N^\prime+1)-1}}(\tau)=(\mbox{polynomial of }T_2)T_{2N^\prime+1}\frac{1}{\Delta}.
\enq
This shows that $N=2^{3M}(2N^\prime+1)-1$ cases are gapful cases.
Indeed several examples were listed up in Sec.4.1.
\paragraph{b) $E_7$}~ \\
By using the theorem we find
\beq
T_8\frac{1}{\Delta}=T_2(T_4-2^{-13}T_1)\frac{1}{\Delta}.
\enq
Thus we can read
\beq
{\tilde \psi}_{E_7}(\tau)=(T_4-2^{-13}T_1)\frac{1}{\Delta}.
\enq
$E_7$ is also gapful case.
\paragraph{c) $E_8$}~\\ We find
\beq
9^{13}T_9\frac{1}{\Delta(\tau)}={\tilde \psi}_{E_8}(\tau)={\tilde Z}_{E_8}(\tau).
\enq
Furthermore the deformed Hecke operator is not necessary for $E_8$ and we find
\beq
Z_{E_8}^\prime(\tau)={\tilde Z}_{E_8}(\tau)=9^{13}T_9\frac{1}{\Delta}.
\enq
This has already been seen in Sec.4.2.

As we have seen in the above discussion,
the condition to obtain the gapful partition functions
depends whether $T_{(rank({\cal G})+1)}$ has the factor $T_{|\Gamma_{\cal G}|}$ or not. When  $T_{(rank({\cal G})+1)}$ does not have $T_{|\Gamma_{\cal G}|}$,
we could not succeed in determining the gapful partition functions in our formalism.

\section{Conclusion and Discussion}
\label{sec:6}
\setcounter{equation}{0}
We determined the twisted ${\cal N}=4~ADE$ partition functions
by using the deformed Hecke operator for some groups.
The resulting partition functions satisfy 
the Montonen-Olive duality and the gap condition at the same time.
All these partition functions were written by
the deformed Hecke operators.
We also discussed the condition of the gapful partition functions
by using the product of the Hecke operators.
We could obtain the gapful partition functions,
when $T_{(rank({\cal G})+1)}$ has the factor $T_{|\Gamma_{\cal G}|}$
as a property of the Hecke operator.

The first remaining problem is to determine
the $ADE$ partition functions not satisfying the condition of the Hecke operators.
There are two cases to consider.
One is the case of $D_{4N+3}(N\ne 2^{3M}(2N^\prime +1)-1)$.
The partition function of this can be written by the deformed Hecke operator, but cannot be deformed into the gapful partition function.
The other is the case of $D_{2N}$ and $E_6$.
The partition function of this cannot be written by the deformed Hecke operator.
In either case, we assume that the total partition functions
may not be written only by the deformed Hecke operators.
We also have to consider the possibility that the partition functions cannot be gapful.\\
The second remaining problem is the interpretation of $\psi_{\cal G}^\prime(\tau)-  \psi_{\cal G}(\tau)$.
This term must have the physical meaning like $\psi_{\cal G}(\tau)$.\\
We also assume that the $ADE$ blow-up formula can be written by the level $|\Gamma_{\cal G}|$ affine Lie algebra.
In our previous results for ${\cal G}=A$ \cite{jin2}, this point was pointed out and we presume that this conjecture can be generalized to $D,E$ cases.\\
Recently Dijkgraaf and Vafa proposed the methods of computing the ${\cal N}=1$
 superpotential as a function of the glueball chiral superfield, such as in \cite{d-v}.
 They argued that this function includes the non-perturbative effects
 such as the Montonen-Olive duality.
 Our results may check their methods in the context of the Montonen-Olive duality. 

{\bf Acknowledgment}\\
We would like to thank M.Jinzenji for helpful suggestions and useful discussions. We also thank Prof.N.Kawamoto for carefully reading our manuscript.

\renewcommand{\theequation}{\alph{section}.\arabic{equation}}
\appendix
\section{Flux Sum}
\label{sec:6}
\setcounter{equation}{0}
For 
$u,v \in H^2(K3,{\bf Z}_N)$ ($N=ad=ap{\tilde d}$),
we verify the following flux sum identity,
\beqy
&&
\sum_{u\in H^2(K3,{\bf Z}_{N})}\zeta_N^{u\cdot v} \delta_{a{\tilde d}u,0}\zeta_{a{\tilde d}}^{-\frac{ab^\prime}{2}(\frac{u}{p})^2}
\enqy
When $a{\tilde d}u=0$, we put $u=p{\tilde u}$.
We perform $u$ summation and obtain,
\beqy
&=&
\sum_{{\tilde u}\in H^2(K3,{\bf Z}_{a{\tilde d}})}\zeta_{a{\tilde d}}^{{\tilde u}\cdot v}\zeta_{a{\tilde d}}^{-\frac{ab^\prime}{2}{\tilde u}^2}
\label{fs1}
\enqy
If $v\ne a{\tilde v}$, we have (\ref{fs1})$=0$. Thus we introduce $\delta_{dv,0}$ and put $v=a{\tilde v}$,
\beqy
&=&
\delta_{dv,0}\sum_{{\tilde u}\in H^2(K3,{\bf Z}_{a{\tilde d}})}\zeta_{a{\tilde d}}^{a{\tilde u}\cdot {\tilde v}}\zeta_{a{\tilde d}}^{-\frac{ab^\prime}{2}{\tilde u}^2}
\no\\&=&
\delta_{dv,0}\sum_{{\tilde u}\in H^2(K3,{\bf Z}_{a{\tilde d}})}\zeta_{{\tilde d}}^{{\tilde u}\cdot {\tilde v}-\frac{b^\prime}{2}{\tilde u}^2}
\enqy
We use $b^\prime {\tilde b}=-1$ mod ${\tilde d}$ and obtain,
\beqy
&=&
\delta_{dv,0}\sum_{{\tilde u}\in H^2(K3,{\bf Z}_{a{\tilde d}})}\zeta_{{\tilde d}}^{-b^\prime{\tilde b}{\tilde u}\cdot {\tilde v}-\frac{b^\prime}{2}{\tilde u}^2}
\no\\&=&
\delta_{dv,0}\zeta_{\tilde d}^{\frac{b^\prime{\tilde b}^2}{2}{\tilde v}^2}\sum_{{\tilde u}\in H^2(K3,{\bf Z}_{a{\tilde d}})}\zeta_{{\tilde d}}^{-\frac{b^\prime}{2}({\tilde u}+{\tilde b}{\tilde v})^2}
\enqy
We perform ${\tilde u}$ summation and finally obtain,
\beqy
&=&
\delta_{dv,0}\zeta_{\tilde d}^{-\frac{{\tilde b}}{2}{\tilde v}^2}(a^2{\tilde d})^{11}
\no\\&=&
\delta_{dv,0}\zeta_{a\tilde d}^{-\frac{a{\tilde b}}{2}{\tilde v}^2}(a^2{\tilde d})^{11}.
\enqy

\end{document}